\documentclass[aps,prd,showpacs,onecolumn,nofootinbib,superscriptaddress,amsmath,amssymb]{revtex4-2}

\usepackage{subcaption}

\usepackage[normalem]{ulem}

\usepackage{soul}
\usepackage{booktabs}
\usepackage{multirow}
\usepackage{microtype}
\usepackage{graphicx}
\usepackage{graphics}
\usepackage{color}
\usepackage{epsfig}
\usepackage{definitions}%
\usepackage{dsfont,amsmath,bm,braket}

\newcommand{\rz}[1]{\textcolor{green}{}}

\newcommand{\rv}[1]{{#1}}

\newcommand{\rvv}[1]{{#1}}
\newcommand{\add}[1]{{#1}}
\newcommand{\new}[1]{{#1}}

\begin{document}
\title{Wigner - Weyl calculus in description of non - dissipative transport phenomena }

\author{M.A. Zubkov }
\email{mikhailzu@ariel.ac.il}
\affiliation{Physics Department, Ariel University, Ariel 40700, Israel}

\date{\today}

\begin{abstract}
Application of Wigner - Weyl calculus to the investigation of non - dissipative transport phenomena is reviewed. We focus on the quantum Hall effect, Chiral Magnetic effect, and Chiral separation effect, and discuss the role of interactions, inhomogeneity, and deviations from equilibrium.
\end{abstract}


\maketitle

\section{Introduction}

Wigner-Weyl calculus originates from the works of  H. Groenewold \cite{Groenewold1946} and J. Moyal \cite{Moyal1949}. It replaces the ordinary  quantum mechanics by the quantum mechanics written using Weyl symbols of operators that are functions on phase space. The ideas realized in Wigner - Weyl calculus were proposed earlier by H. Weyl \cite{Weyl1927} and E. Wigner \cite{Wigner1932} themselves. In this formalism Weyl symbols are used  instead of the the operators of physical quantities. The product of two operators is replaced by the Moyal product of two functions  \cite{Ali2005,Berezin1972}. This calculus has been applied to the solution of quantum mechanical problems \cite{Curtright2012,Zachos2005}. The most well - known problem of this type is the unharmonic oscillator. Within the quantum  field theory the Wigner-Weyl formalism was applied to several problems of high energy physics theory and to condensed matter physics \cite{Cohen1966,Agarwal1970,E.C.1963,Glauber1963,Husimi1940,Cahill1969,Buot2009}.
For the applications to QCD see \cite{Lorce2011,Elze1986}. There were also numerous applications of this calculus to  quantum kinetic theory  \cite{Hebenstreit2010,Calzetta1988}.

It is worth mentioning that it is not easy to extend Wigner-Weyl calculus to lattice models, although the attempts to built such a formalism started long time ago   \cite{Buot1974,Buot2009,Buot2013}, see also  \cite{WOOTTERS19871,Leonhardt1995,KASPERKOVITZ199421,Ligabo2016}. The so - called approximate version of the lattice Wigner-Weyl calculus has been proposed in  the participation of the present author \cite{ZW2019}.
This calculus is approximate because its  application is limited to the systems with weak inhomogeneity, which means, in particular, that the external magnetic field strength should be much smaller than $10000$ Tesla. In practise this requirement is always fulfilled in real solid state systems and in the lattice regularized relativistic quantum field theory. It appears that within this formalism one may express through the topological invariants the response of various nondissipative currents to external field strength \cite{Zubkov2017,Chernodub2017,Khaidukov2017,Zubkov2018a,Zubkov2016a,Zubkov2016b,Chernodub2016}.

The topological invariants  considered in \cite{Zubkov2017,Chernodub2017,Khaidukov2017,Zubkov2018a,Zubkov2016a,Zubkov2016b,Chernodub2016} are composed of the Green funcitons. The invariants of this kind were considered much more earlier. The simplest ones are responsible for the stability of the Fermi surface in the $3+1$ D systems:
\be
N_1= \tr \oint_C \frac{1}{2\pi \ii}
G(p_0,\bp)d G^{-1}(p_0,\bp).
\ee
Here $C$ is an arbitrary contour, which encloses the Fermi surface \cite{Volovik2003a} within momentum space.
The topological stability of Fermi points is protected by \cite{Matsuyama1987a,Volovik2003a}
\be
N_3=\frac1{24\pi^2} \epsilon_{\mu\nu\rho\lambda} \tr\int_S dS^\mu
G\partial^\nu G^{-1}
G\partial^\rho G^{-1}
G\partial^\lambda G^{-1}.
\ee
Here $S$ is the surface that surrounds the Fermi points.

The first topological expression for the  Hall conductivity was given through the TKNN invariant \cite{Thouless1982}. This invariant has been defined for the systems in the presence of constant external magnetic field, and for the description of intrinsic anomalous quantum Hall effect in homogeneous systems \cite{Avron1983,Fradkin1991,Tong:2016kpv,Hatsugai1997,Qi2008}. Strictly speaking, the TKNN invariant may be applied only to the systems with constant magnetic field or homogeneous quantum Hall insulators. In the presence of interactions the TKNN invariant is not defined.

In the absence of interactions the TKNN invariant for the intrinsic QHE (existing without external magnetic field) may be re - written through the momentum space Green's function \cite{IshikawaMatsuyama1986,Volovik1988} (see also Chapter 21.2.1 in \cite{Volovik2003a}).
For the $2+1$D fermions  the Hall conductivity is given by
$$
\sigma_H = \frac{\cal N}{2\pi},
$$
where
{\be
{\cal N}
=  \frac{ \epsilon_{ijk}}{  \,3!\,4\pi^2}\, \int d^3p \Tr
\[
{G}(p ) \frac{\partial {G}^{-1}(p )}{\partial p_i}  \frac{\partial  {G}(p )}{\partial p_j}  \frac{\partial  {G}^{-1}(p )}{\partial p_k}
\].
\label{calM2d230}
\ee}%
In the presence of interactions this  expression remains valid, if the non-interacting two-point Green's  function has been substituted by the Green's function with the interaction corrections \cite{ColemanHill1985,Lee1986,ZZ2019}. The original TKNN invariant cannot be extended in a similar way to the interacting systems in principle.

In the present paper we review the recent works  by the group working in Ariel University on the topological description of non - dissipative transport in non - homogeneous systems. We summarize here the results reported earlier in a series of papers (see  \cite{ZW2019,ZZ2019_2,FSWZZ2020,ZZ2021,SZ2020,BLZ2021,BFLZZ2021,ZZ2019_FeynRule} and references therein). We used Wigner - Weyl formalism and start from the extension of Eq. (\ref{calM2d230}) to the systems with weak inhomogeneity. In particular, such an inhomogeneity may be caused by disorder (in case of the solid state systems), or non - uniform external fields. It appears that the QHE conductivity remains topological even in this case. Moreover, we consider the perturbative corrections due to interactions. We will see that the QHE conductance remains robust to these corrections.

Another non - dissipative transport effect that will be considered in the present paper is the chiral magnetic effect (CME) \cite{Vilenkin,CME,Kharzeev:2013ffa,Kharzeev:2009pj,SonYamamoto2012}. It is widely believed that it does not appear in true equilibrium \cite{Valgushev:2015pjn,Buividovich:2015ara,Buividovich:2014dha,Buividovich:2013hza,Z2016_1,Z2016_2,nogo,nogo2}. It might appear out of equilibrium in a steady state with the chiral imbalance due to external electric field and external magnetic field \cite{Nielsen:1983rb}. Experimental detection of this effect was reported through the study of magnetoresistance in Dirac and Weyl semimetals \cite{ZrTe5}. In the absence of external electric field the chiral magnetic effect was predicted for  the non - equilibrium systems with dependence on time in chiral chemical potential  \cite{Wu:2016dam}. In the present paper we concern the setup of equilibrium CME in the presence of inhomogeneity, finite temperature and interactions. We will see that under these circumstances the CME still does not exist in equilibrium.

The chiral separation effect (CSE) was proposed in \cite{Metl}. This is the  appearance of an axial current in the direction of external magnetic field. In the chiral limit the axial current is proportional to the external magnetic field  $B^k$ and the ordinary chemical potential $\mu$ counted from the Fermi point:
\begin{equation}
	{\bf J}_5 = \frac{1}{2\pi^2}  \mu {\bf B}\label{1}
\end{equation}
It was proposed to observe this effect in experiment during the heavy ion collisions \cite{Kharzeev:2015znc,Kharzeev:2009mf,Kharzeev:2013ffa,ref:HIC}. The quark  - gluon plasma  \cite{QCDphases,1,2,3,4,5,6,7,8,9,10}, presumably, appears within the fireballs resulted from the collisions of two heavy ions. Quarks in this state of matter are  massless (if their current mass is neglected). Therefore, we deal with the chiral fermions.  External magnetic field exists because the motion of ions creates  electric currents. The non - central collisions produce strong magnetic field orthogonal to the trajectories of colliding ions. The CSE gives axial current within the fireball, which may be detected after the decay of the fireball indirectly through the distribution of outgoing particles. Relation between the CSE and  the chiral anomaly has been discussed, for example,  in \cite{Zyuzin:2012tv}.
The detailed consideration of the CSE may be found, for example, in  \cite{Gorbar:2015wya,Buividovich:2013hza,KZ2017}.  Interaction corrections to CSE have been considered, for example, in  \cite{Shovkovy}. The CSE was considered mostly for the homogeneous systems, while in any real experiments we deal with the inhomogeneous systems. Here we will concentrate on this situation and describe the CSE for the nonhomogeneous systems.

In order to extend our consideration to the systems out of equilibrium we use  Keldysh formalism \cite{Keldysh64,Bonitz00,BS03,BF06,KB62,Baym62,Schwinger61}. The finite temperature equilibrium quantum statistical mechanics \cite{Matsubara55,BdD59,Gaudin60,AGD63}  and the real time QFT represent the limiting cases of Keldysh QFT. The functional integral formalism \cite{Kamenev} as well as the operator formalism \cite{Langreth76,Rammer07} are used in Keldysh technique as well as in the equilibrium theory. The perturbation theory of Keldysh formalism is technically similar to the one of the conventional QFT \cite{Keldysh64,Langreth76}. Schwinger-Dyson equations being applied to Keldysh technique \cite{Landau56,LW60,Luttinger60,Baym62} provide the natural framework for the calculation of transport properties \cite{CCNS71,AG75,LO75,Bonitz00,BS03,BF06,Langreth76,Danielewicz84,CSHY85,RS86,Rammer07}.
Keldysh formalism has been applied to QCD \cite{CGC}.
A version of Wigner-Weyl calculus similar to the one of equilibrium theory \cite{ZZ2019_FeynRule,ZW2019} has been developed within the Keldysh formalism \cite{Shitade, Sugimoto,Sugimoto2006,Sugimoto2007,Sugimoto2008}. The gradient expansion together with the Schwwinger - Dyson equations here allows to calculate fermion propagator.  In \cite{Mokrousov,Mokrousov2} the  development of this technique was given  that allows to deal with the non - homogeneous systems. In the present paper we will discuss the further development of this line of research, and consider the application of Wigner - Weyl calculus to the non - dissipative transport phenomena for the systems out of equilibrium using an example of QHE.

The paper is organized as follows. In Sect. \ref{SectII} we recall the basic notions of Wigner-Weyl calculus within the inhomogeneous theory at zero temperature. We follow here the approach of \cite{ZW2019}. In Sect. \ref{CMEQHE} we consider the QHE at zero temperature \cite{ZW2019} and equilibrium CME at finite temperature \cite{BLZ2021}. In Sect. \ref{SectIII} we consider chiral separation effect and review results of \cite{SZ2020}. In Sect. \ref{SectIV} we discuss loop corrections to the QHE conductivity of the two - dimensional non  - homogeneous systems. We summarize here the results reported in \cite{ZZ2019_2} and \cite{ZZ2021}. In Sect. \ref{SectKeldysh} we briefly summarize the results of \cite{BFLZZ2021} on the QHE out of equilibrium.
 In Sect. \ref{SectConcl} we end with the conclusions.

\section{Wigner - Weyl calculus in equilibrium inhomogeneous theory. }

\label{SectII}

\subsection{Approximate Wigner - Weyl calculus}

Here we closely follow \cite{ZW2019}.
We consider the lattice model with the partition function in momentum space at very small temperatures \cite{Z2016}
\begin{eqnarray}
	Z &=& \int D\bar{\psi}D\psi \, {\rm exp}\Big(  \int_{\cal M} \frac{d^D {p}_1}{\sqrt{|{\cal M}|}} \int_{\cal M} \frac{d^D {p}_2}{\sqrt{|{\cal M}|}}\bar{\psi}^T({p}_1){\cal Q}(p_1,p_2)\psi({p}_2) \Big),\label{Z1}
\end{eqnarray}
Here $D$ is dimension of space - time. It is equal to $3$ if we are considering the two - dimensional systems, and $4$ if we are considering the three dimensional ones. Notice that we discretize imaginary time in the same way as spatial coordinates. For the condensed matter systems the lattice spacing in imaginary time direction is to be set to zero at the end of calculations, while the lattice spacing in spatial direction remains equal to the real distance between the adjacent lattice sites. For the lattice regularized relativistic quantum field theory the common lattice spacing is to be send to zero considering the continuum limit.
By $|{\cal M}|$ we denote volume of $D$ -  dimensional momentum space $\cal M$. Both $p_1$ and $p_2$ are $D$ - momenta. Here matrix elements of lattice Dirac operator $\hat{Q}$ are denoted by
$$
\langle p_1| \hat{Q} | p_2 \rangle = {\cal Q}(p_1,p_2)
$$
We use the units, in which both $\hbar $ and $c$ are equal to unity.  Elementary charge of electron $e$ is included to the definition of electric and magnetic fields.

Wigner transformation of Green function ${\cal G}(p_1,p_2) = \langle p_1 | \hat{G} | p_2\rangle$ that is the Weyl symbol of operator $\hat{G}$ is given by
\begin{equation}
	{G}_W(x,p) \equiv \int_{\cal M} dq e^{ix q} {\cal G}({p+q/2}, {p-q/2})\label{GWx}
\end{equation}
Weyl symbol of operator $\hat{Q}$ is:
$$
{Q}_W(x,p) \equiv \int_{\cal M} dq e^{ix q} {\cal Q}({p+q/2}, {p-q/2})$$
We assume here that the inhomogeneity is weak, i.e.  variations of external fields (including disorder potential) at the distance of the order of lattice spacing may be neglected.
In this situation matrix element $Q(p_1,p_2)$ remains nonzero only when $|p_1-p_2|$ is much smaller than the size of momentum space. Under these conditions  Wigner transformed Green function and Weyl symbol of Dirac operator satisfy  Groenewold equation
\begin{equation}
	{G}_W(x,p) \star Q_W(x,p) = 1 \label{Geq}
\end{equation}
Here the Moyal product is introduced
\begin{equation}
	\star \equiv
	e^{\frac{i}{2} \left( \overleftarrow{\partial}_{x}\overrightarrow{\partial_p}-\overleftarrow{\partial_p}\overrightarrow{\partial}_{x}\right )}
	\label{GQW}
\end{equation}
Notice that the Moyal product is associative. Therefore, we remove brackets inside our expressions. Also we are using the following properties of Moyal product:
$$
\int d^D x d^D p {\rm Tr} A_W \star B_W = \int d^D x d^D p {\rm Tr} A_W  B_W
$$

External electromagnetic field gives rise to magnetic field ${\bf{B}}({\bf{x}})=\nabla\times {\bf{A}}({\bf{x}})$ and electric field ${\bf E}({\bf x}) = - \nabla \phi({\bf x})$. In our Euclidean space $A_{D+1} = - i \phi$, while spatial components $A_k = A^k$ with $k=1,..., D-1$. $Q$ depends on the combination $\pi = p-A(x)$:
$Q_W(p,x) = Q(p-A(x),x)$. Here $Q(p,x)$ is Weyl symbol of operator $\hat{Q}$ in the absence of external electromagnetic field $A$. In the following we will be writing sometimes $Q_W(p-A(x),x)$.

\subsection{Linear response of Green function to external electric/magnetic field}

The Dirac operator may be written as
$$
Q_W(p,x)\approx Q_W^{(0)}(p,x)+\delta Q_W(p,x)
$$
where $\delta Q_W=-{\partial_p}_kQ^{(0)}(p,x)A_k$ while $Q^{(0)}$ is Dirac operator with $A=0$. The Green function may be represented as
$$
G_W(p,x)\approx G_W^{(0)}(p,x)+\delta G_W(p,x).
$$
The Groenewold equation reads
\begin{eqnarray}
	&&\left(Q_W^{(0)}(p,x)+\delta Q_W(p,x)\right)\star\left(G_W^{(0)}(p,x)+\delta G_W(p,x)\right)\\\nonumber&=& Q^{(0)}_W\star G^{(0)}_W+Q^{(0)}_W\star \delta G_W+\delta Q_W\star G^{(0)}_W=1,
\end{eqnarray}
with $Q^{(0)}_W\star G^{(0)}_W=1$. We denote the field strength by  $F_{ij}=\partial_iA_j-\partial_jA_i$. Hence
\begin{eqnarray}
	\delta G_W&=&G^{(1)}_{W(k)}A_k+G^{(2)}_{W(ij)}F_{ij}\nonumber\\&=&\left[G^{(0)}_W\star{\partial_p}_kQ^{(0)}_W(p,x)\star G^{(0)}_W\right]A_k+\frac{i}{2}\left[G^{(0)}_W\star{\partial_p}_iQ^{(0)}_W(p,x)\star G^{(0)}_W\star{\partial_p}_jQ^{(0)}_W(p,x)\star G^{(0)}_W\right]F_{ij}\nonumber.
\end{eqnarray}
Thus we have
$$
G_W(p,x) \approx G^{(0)}_W(p,x)+G^{(1)}_{W(k)}A_k+G^{(2)}_{W(ij)}F_{ij}
$$
where
$$
G^{(1)}_{W(k)}=G^{(0)}_W\star{\partial_p}_kQ^{(0)}_W(p,x)\star G^{(0)}_W
$$
$$
G^{(2)}_{W(ij)}=\frac{\ii}{2}\left[G^{(0)}_W\star{\partial_p}_iQ^{(0)}_W(p,x)\star G^{(0)}_W\star{\partial_p}_jQ^{(0)}_W(p,x)\star G^{(0)}_W\right]
$$

\section{Equilibrium CME and QHE}
\label{CMEQHE}
\subsection{Zero temperature}

Electric current may be expressed as
$$
j_k(x)=\frac{\delta \rm ln Z}{\delta A_k(x)}=-\frac{1}{2}\int_{\mathcal M}\frac{d^Dp}{(2\pi)^D}\rm tr\left[G_W(p,x) {\partial_p}_kQ_W(p,x)\right]+c.c
$$
By $J$ we denote the electric current  integrated over the whole (Euclidean) space - time. Integration over imaginary time automatically gives factor $1/T$, where $T$ is temperature that is assumed to be small.
$$
J_k=\int dx j_k(x)=-\frac{1}{2} \int d^{D}x\int_{\mathcal M}\frac{d^Dp}{(2\pi)^D}\rm tr\left[G_W(p,x) {\partial_p}_kQ_W(p,x)\right]+c.c.
$$
Here and below we replace the sum over the lattice sites by an integral. This is possible because we assume weak inhomogeneity of the considered systems.
We represent  ${\partial_p}_kQ_W(p,x)={\partial_p}_kQ^{(0)}_W(p,x)-\left({\partial_p}_k{\partial_p}_jQ^{(0)}_W(p,x)\right)A_j.$ Therefore,
$$
G_W(x,p){\partial_p}_kQ_W(p,x)=\left(G^{(0)}_W(p,x)+G^{(1)}_{W(l)}A_l+G^{(2)}_{W(mn)}F_{mn}\right)\left({\partial_p}_kQ^{(0)}_W(p,x)-\left({\partial_p}_i
{\partial_p}_jQ^{(0)}_W(p,x)\right)A_j\right).
$$
Current density is given by
\begin{eqnarray}
	j_i(x)&=&-\int_{\mathcal M}\frac{d^Dp}{(2\pi)^D}\rm Tr\Large[G^{(0)}_W{\partial_p}_iQ^{(0)}_W-{\partial_p}_k\left(G^{(0)}_W{\partial_p}_iQ^{(0)}_W\right)A_k\nonumber\\&-&\frac{i}{2}\left[G^{(0)}_W\star{\partial_p}_mQ^{(0)}_W(p,x)\star G^{(0)}_W\star{\partial_p}_nQ^{(0)}_W(p,x)\star G^{(0)}_W\right]{\partial_p}_iQ^{(0)}_W F_{mn}\Large].
\end{eqnarray}
We represent
$$
j_i(x)=j_i^{(0)}(x)+j_{i(k)}^{(1)}(x)A_k(x)+j_{i(mn)}^{(2)}(x)F_{mn}(x)
$$
with
$$
j_i^{(0)}(x)=-\int_{\mathcal M}\frac{d^Dp}{(2\pi)^D}\rm Tr\left[G^{(0)}_W{\partial_p}_iQ^{(0)}_W\right] ~~\text{and}~~
j_{i(k)}^{(1)}(x)=\int_{\mathcal M}\frac{d^Dp}{(2\pi)^D}\rm Tr
\left[{\partial_p}_k\left(G^{(0)}_W{\partial_p}_iQ^{(0)}_W\right)\right].$$
$j_{i(k)}^{(1)}(x)=0$ as an integral of total derivative. We arrive at
$$
j_{i(mn)}^{(2)}(x)=-\frac{i}{2}\int_{\mathcal M}\frac{d^Dp}{(2\pi)^D}\rm Tr\left[\left[G^{(0)}_W\star{\partial_p}_mQ^{(0)}_W(p,x)\star G^{(0)}_W\star{\partial_p}_nQ^{(0)}_W(p,x)\star G^{(0)}_W\right]{\partial_p}_iQ^{(0)}_W\right].
$$
The term in the average current that is linear in the field strength $F_{mn}$ is
\begin{eqnarray}
	\bar{J}_i^{(2)}&=&\frac{J_i^{(2)}}{V^{(D)}}\equiv \frac{F_{mn}}{V^{(D)}}\int d^Dx j^{(2)}_{i(mn)}(x)=\frac{T F_{mn}}{V}\int dx j^{(2)}_{i(mn)}(x)\\\nonumber&=&\frac{-iF_{mn}}{2V}\int d^Dx \int_{\mathcal M}\frac{d^Dp}{(2\pi)^D}\rm Tr\left[\left[G^{(0)}_W\star{\partial_p}_mQ^{(0)}_W(p,x)\star G^{(0)}_W\star{\partial_p}_nQ^{(0)}_W(p,x)\star G^{(0)}_W\right]{\partial_p}_iQ^{(0)}_W\right],
\end{eqnarray}
Here $V^{(D)}$ is volume of Euclidean space-time while $V$ is spatial volume, $T$ is temperature that tends to zero.
For $D=4$ we obtain expression for the average current  \cite{ZW2019}
$$
\bar{J}^k=\frac{1}{4\pi^2}\epsilon^{ijkl}\mathcal{M}_lF_{ij}
$$
where
$$
\mathcal{M}_l=\frac{-iT\epsilon_{ijkl}}{3!V8\pi^2}\int d^Dx \int_{\mathcal M}{d^Dp}\rm Tr\left[\left[G^{(0)}_W\star{\partial_p}_iQ^{(0)}_W(p,x)\star G^{(0)}_W\star{\partial_p}_jQ^{(0)}_W(p,x)\star G^{(0)}_W\right]{\partial_p}_kQ^{(0)}_W\right]
$$
In this expression the star may be inserted:
$$
\mathcal{M}_l=\frac{-iT\epsilon_{ijkl}}{3!V8\pi^2}\int d^Dx \int_{\mathcal M}{d^Dp}\rm Tr\left[G^{(0)}_W\star{\partial_p}_iQ^{(0)}_W(p,x)\star G^{(0)}_W\star{\partial_p}_jQ^{(0)}_W(p,x)\star G^{(0)}_W\star {\partial_p}_kQ^{(0)}_W\right]
$$
The last expression is topological invariant if there is no singularity inside the integral. If we consider the magnetic field only, it is calculated through the field tensor $F_{ij}=\epsilon_{ijk}B_k$. Notice, that we define the field strength in Euclidean space - time, and assume that $A^k = A_k$. Therefore, from ${\bf B} = {\rm rot}\, {\bf A}$ it follows $B_k = \epsilon_{kij} \partial_i A^j$. {The absence of Chiral Magnetic Effect in equilibrium non - homogeneous system follows from the fact that in the presence of finite chiral chemical potential the system remains gapped. This property remains valid, at least, as long as the inhomogeneity is sufficiently weak. As a result the topological nature of quantity $M_4$ means that its response to the variation of chiral chemical potential is vanishing, This means the vanishing of the so - called CME conductivity.} This result has been obtained in \cite{BLZ2021}, and it extends  the result of \cite{Z2016} to the non - homogeneous systems \footnote{It is worth mentioning that in \cite{Z2016} there was a mistake (wrong sign) in the second row of Eq. (8), consequently the signs are to be changed to inverse in Eqs. (10), (11), (32), (33), (34), (35), (38), the second and the third rows in Eq. (31) of \cite{Z2016};   see also corrigendum in \cite{ZW2019}. Fortunately, this wrong sign did not affect the physical results of \cite{Z2016}. }.

The similar consideration gives for the case of the two - dimensional system (D=3):
$$
\bar{J}^k=\frac{1}{4\pi}\epsilon^{ijk}\mathcal{M} F_{ij}
$$
where
$$
\mathcal{M}=-\frac{iT\epsilon_{ijk}}{3!V 4\pi^2}\int d^Dx \int_{\mathcal M}{d^Dp}\rm Tr\left[G^{(0)}_W\star{\partial_p}_iQ^{(0)}_W(p,x)\star G^{(0)}_W\star{\partial_p}_jQ^{(0)}_W(p,x)\star G^{(0)}_W\star {\partial_p}_kQ^{(0)}_W\right]
$$
The last expression is the topological invariant again.

If we consider the external electric field $E_k$ originated from electromagnetic potential $A_k$, then the nonzero components of Euclidean field strength $F_{ij}$ are given by
$F_{Dk} = \partial_D A_k - \partial_k A_D = -i E_k$.  Then we arrive at
\begin{equation}
\bar{J}^k=\frac{1}{2\pi}\epsilon^{kj}\mathcal{N} E_{j}\label{JQHE}
\end{equation}
where
\begin{equation}
\mathcal{N}=\frac{T\epsilon_{ijk}}{3!V 4\pi^2}\int d^Dx \int_{\mathcal M}{d^Dp}\rm Tr\left[G^{(0)}_W\star{\partial_p}_iQ^{(0)}_W(p,x)\star G^{(0)}_W\star{\partial_p}_jQ^{(0)}_W(p,x)\star G^{(0)}_W\star {\partial_p}_kQ^{(0)}_W\right]\label{JQHE2}
\end{equation}
{\bf This is the topological expression for the QHE conductivity of non - homogeneous system.} It has been proposed for the first time in \cite{ZW2019}.

\subsection{Theory at nonzero temperature. Chiral magnetic conductivity.}

At finite temperature
the above expression for the partition function is to be modified. It receives the form
\begin{eqnarray}
	Z &=& \int D\bar{\psi}D\psi \, {\rm exp}\Big( 2\pi T  \sum_{p_1^4=p_2^4=\omega}\int \frac{d^{D-1} {p}_1}{\sqrt{|{\cal M}|}} \int \frac{d^{D-1} {p}_2}{\sqrt{|{\cal M}|}}\bar{\psi}^T({p}_1){\cal Q}(p_1,p_2)\psi({p}_2) \Big),\label{Z1}
\end{eqnarray}
Here $\sum_{\omega}$ is the sum over Matsubara frequencies $\omega = 2 \pi T (n+1/2)$, $n \in Z$, $0\le n < N$, while $N = 1/T$ is the number of lattice sites in the fourth direction. Temperature $T$ in this expression is measured in the lattice units. Integral $\int \frac{d^{D-1} {p}_1}{\sqrt{|{\cal M}|}}$  is over the Brillouin zone $\cal B$. Here ${\cal M}=2\pi |{\cal B}|$, where $|{\cal B}|$ is volume of the Brillouin zone.

For $D=4$ the same steps as above lead us to the following expression for the response of  electric current to spatial components of external field srength
$$
\bar{J}^k=\frac{1}{4\pi^2}\epsilon^{ijk4}\mathcal{M}_4F_{ij}.
$$
with
$$
\mathcal{M}_4 = 2\pi T \sum\limits_{\omega} \mathcal{N}_4(\omega)
$$
and
\begin{equation}
	\mathcal{N}_4(\omega)=\frac{- i \epsilon_{ijk4}}{3!V8\pi^2}\int d^{D-1} x \int_{\mathcal B}{d^{D-1}p} \rm Tr\left[G^{(0)}_W\star{\partial_p}_iQ^{(0)}_W(p,x)\star G^{(0)}_W\star{\partial_p}_jQ^{(0)}_W(p,x)\star G^{(0)}_W\star{\partial_p}_kQ^{(0)}_W\right].
\end{equation}
Notice that both $Q_W$ and $G_W$ do not depend  on $\tau$. As a result the Moyal product does not contain derivatives with respect to $\omega$ and $\tau$. For any value of Matsubara frequency $\omega$ the expression for  $\mathcal{N}_4(\omega)$ is  topological invariant. Finite temperature provides that there is no singularity inside the corresponding integral over momenta. As a result the response of $\mathcal{M}_4$ to variation of any parameter is zero. {\bf Taking the chiral chemical potential at such a parameter, we come to the absence of chiral magnetic effect at finite temperature for the non - homogeneous systems, at least, if the inhomogeneity is sufficiently weak.}

\section{Axial current and chiral separation effect}

\label{SectIII}
\subsection{Axial current}

Here we consider the chiral separation effect, which is the  appearance of axial current in the $4D$ fermionic systems with finite chemical potential and external magnetic field. Here we follow \cite{SZ2020}. We assume that the magnetic field is uniform. At the same time the system itself contains weak inhomogeneity. The Dirac operator $\hat Q$ is a  $4\times 4 $ matrix with spinor indices.
The axial current may be defined as
\be
j^5_k(x)= -
\int_{\mathcal M} \frac{d^Dp}{|\mathcal M|}
\tr \left[ \gamma^5 G_W(x,p) \pd_{p_k} Q_W(x,p)  \right]
\label{ji5x}\ee
Now we can repeat all steps considered above for electric current, and obtain the  response of axial current to external field strength:
\be
j_k^5(x)&=-
\frac{i}{2}
\int_{\mathcal M} \frac{d^Dp}{|\mathcal M|}
\tr \Bigl[\gamma^5
\Big[ G_W^{(0)}\star \(\pd_{p_i} Q_W^{(0)}\) \\&\star G_W^{(0)}
\star \(\pd_{p_j} Q_W^{(0)}\) \star G_W^{(0)} \Big] \pd_{p_k}Q_W^{(0)}
\Bigr]
F_{ij}
\label{ji5lr}\ee
The axial current averaged over the system volume is given by
\be
&\bar{J}_k^5=
-\frac{i}{2}\frac{1}{\beta{\bf V}}
\int d^Dx\int_{\mathcal M} \frac{d^Dp}{(2\pi)^D}\tr \Bigl[\gamma^5
\Bigl[ G_W^{(0)}\star \(\pd_{p_i} Q_W^{(0)}\) \\&\star G_W^{(0)}
\star \(\pd_{p_j} Q_W^{(0)}\) \star G_W^{(0)} \Bigr] \pd_{p_k}Q_W^{(0)}
\Bigr]
F_{ij}
\label{Ji5lr}\ee

We regularize the theory by finite (but small) temperature with Matsubara frequencies
$
p_4=\omega_n=\frac{2\pi\(n+\frac{1}{2}\)}{\beta}
$.
As above, inverse temperature $\beta = 1/T$ is taken in lattice units:
$
N_t\equiv\frac{1}{T}
$.
 $\om_n$ never equals to zero, and, therefore, even for the system of gapless fermions the propagator does not have poles in momentum space.
Introducing the chemical potential $\omega_n\ra\omega_n-i\mu$
we obtain the following response to magnetic field and chemical potential
\be
&\bar{J}_k^5=-
\frac{1}{{2\bf V}\beta}
\sum_{n=0}^{{N_t}-1}
\int d^3x \int_{\mathcal B} \frac{d^3p}{(2\pi)^3}\pd_{\omega_n}\tr \Bigg[\gamma^5
\Big[ G_W^{(0)}\star \(\pd_{p_i} Q_W^{(0)}\) \star G_W^{(0)}
\star \(\pd_{p_j} Q_W^{(0)}\) \star G_W^{(0)} \Big] \pd_{p_k}Q_W^{(0)}
\Bigg]
F_{ij}\mu
\label{Ji5mu1}\ee
Here $|\mathcal B|$ is volume of the three - dimensional Brillouin zone. We suppose that the external field strength corresponds to a constant magnetic field $B$: $F_{ij} = \epsilon_{ijk} B_k$. Then
$$
\bar{J}_k^5(x)=\mathcal{\sigma}_{ijk}\epsilon_{ijk^\prime} B_{k^\prime}\mu
$$
For the wide range of systems $\epsilon_{ijk} \sigma_{ijk^\prime} = \delta^{k k^\prime} \sigma_{CSE}$ with a scalar CSE conductivity.

\subsection{Topological expression for the CSE conductivity}

We represent expression for the CSE conductivity as
\be
\mathcal{\sigma}_{ijk}=&-
\sum_{n=0}^{{N_t}-1}
\pd_{\om_n}\mathcal{\sigma}_{ijk}^{(3)}
\label{Nijk5_1}\ee
where
\be
\mathcal{\sigma}_{ijk}^{(3)}&=
\frac{1}{{2\bf V}}
\int d^3x \int_{\mathcal B} \frac{d^3p}{(2\pi)^3}\tr \Bigg[\gamma^5
\Big[ G_W^{(0)}\star \(\pd_{p_{[i}} Q_W^{(0)}\) \star G_W^{(0)}
\star \(\pd_{p_{j]}} Q_W^{(0)}\) \star G_W^{(0)} \Big] \pd_{p_k}Q_W^{(0)}
\Bigg]
\label{N_3}\ee

Now we consider the limit of small temperature $T\ra 0$, $N_t\ra \infty$, which means $\frac{\pi}{N_t}=\ep\ra 0$. As a result we replace the sum by an integral, but the point $\omega = 0$ is excluded from this integral because it contains singularity for the system with gapless fermions:
\be
\sum_{n=0}^{{N_t}-1} \tab\ra \tab
 \frac{\beta}{2\pi}\int_{0+\ep}^{2\pi-\ep}d\omega
\ee
We obtain
\be
&\mathcal{\sigma}_{ijk}=
\lim_{\ep\ra0}\[
\mathcal{\sigma}_{ijk}^{(3)}(0+\ep) +\(-\mathcal{\sigma}_{ijk}^{(3)}(0-\ep)\) \]
\label{Nijk_N_3}\\
\ee
with
\begin{widetext}
	\be
	&{\sigma}_{ijk}^{(3)}(\om=0\pm \ep)=&\frac{1}{{2\bf V}}
	\int_{\mathcal B} \frac{d^3p}{(2\pi)^4}
	\int d^3x
	\tr \Bigg[\gamma^5
	\Big[ G_W^{(0)}\star \(\pd_{p_{[i}} Q_W^{(0)}\) \star G_W^{(0)}
	\star \(\pd_{p_{j]}} Q_W^{(0)}\) \star G_W^{(0)} \Big] \pd_{p_k}Q_W^{(0)}
	\Bigg]\Bigg|_{\om=0\pm \ep}
	\label{N_3_1}\ee
\end{widetext}

We discuss the static case, when both  $G$ and $Q$ do not depend on (imaginary) time. In this case all possible singularities of the above expressions may be situated only at $\omega = 0$. The above expressions avoid these singularities because of nonzero $\epsilon$. Notice, that for the homogeneous systems the position of singularity coincides with Fermi surfaces.
Weak inhomogeneity modifies the position of the singularities only slightly.

In Eq. (\ref{Nijk_N_3}) the two integrals cancel each other everywhere in the Brillouin zone except for the small vicinities of the singularities of the function standing inside the integral of Eq. (\ref{N_3_1}). As a result we restrict integrations in Eq. (\ref{N_3_1}) to those  small regions of the Brillouin zone. {\bf In this region we assume the presence of precise chiral symmetry, i.e. in these regions $\gamma^5$ commutes or anti - commutes with the Green function.}
The sum of the integrals in Eq. (\ref{Nijk_N_3}) is a topological invariant. Therefore, we are able to deform the two pieces of the surface above/below the singularity in such a way that the resulting $3D$ closed hypersurface  surrounds the singularities. As a result  instead of the two pieces of the infinitely close planes for any $x$ we integrate over the hypersurface $\Sigma_3(x)$ surrounding the positions of the singularities, and
\begin{equation}
\sigma_{ijk} = \epsilon_{ijk} \sigma_{CSE}/2, \quad	\sigma_{CSE} = \frac{\mathcal{N}}{2\pi^2}\label{sigmaH}
\end{equation}
with
\begin{widetext}
	\begin{eqnarray}
		\mathcal{N}&=&\frac{1}{48 \pi^2 {\bf V}}
		\int d^3x
		\int_{\Sigma_3(x)}
		\tr \Bigg[\gamma^5
		G_W^{(0)}\star d Q_W^{(0)} \star G_W^{(0)}
		\wedge \star d Q_W^{(0)}\star G_W^{(0)} \star \wedge d Q_W^{(0)}
		\Bigg]\label{Ncompl}
	\end{eqnarray}
\end{widetext}
The last expression is topological invariant if $\gamma^5$ commutes or anti - commutes with $Q_W$ and $G_W$ in the vicinity of $\Sigma_3(x)$ for any $x$. The obtained result means that the CSE conductivity is a topological quantity for the nonhomogeneous systems provided that the system is chiral, i.e. commutator/anti - commutator of matrix $\gamma_5$ with $Q_W$ may be neglected in a vicinity of the singularities of expression standing under the integral over momentum for any $x$. If such a system is obtained by a smooth modification of the uniform system with massless Dirac fermioms, the value of $\cal N$ is equal to the number of the species of these Dirac fermions. In the other words, $\cal N$ counts the number of the chiral Dirac fermions in the given system.

\section{(The absence of) loop corrections to the QHE conductivity}
\label{SectIV}

\subsection{Setup of the {\it Gedankenexperiment}. $2+1$ D tight - binding model in the presence of interactions. }

\label{SectCoulomb}


\begin{figure}[h]
	\centering  %
	\includegraphics[width=5cm]{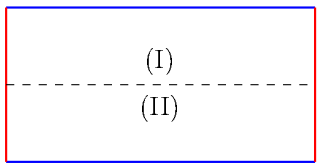} 
	\caption{The two-dimensional fermion system is located on the torus.
		The opposite edges are glued
		together. The torus is divided into the two pieces with opposite directions of external electric field.}  %
	\label{fig_torus}   %
\end{figure}

In this section we consider interaction  corrections to Hall conductivity of the two dimensional systems. We follow closely discussion of \cite{ZZ2019_2}. We deal with the  $2+1$ D tight-binding model with
the interactions between electrons. The system is assumed to be placed at the surface of torus. In the other words, the periodic boundary conditions are assumed (see Fig.\ref{fig_torus} ). The x - axis is along the circle that is placed inside the torus, while the y-axis is along the circle winding along the first circle. We suppose that $y\in (-L,L]$,and  $L$ is assumed very large.

 We  divide the cylinder into the two pieces (Fig.\ref{fig_torus}  ):
	(I) For $y\in [0,L]$ the coupling constant $\alpha \ne 0$, and there are interactions between electrons;
	(II) For $y\in (-L,0)$ the value of coupling constant $\alpha^\prime$ is close to zero $\alpha^\prime \to 0$, and the inter - electron interactions are turned off.

Magnetic field is orthogonal to the surface of the torus.  Its value, however, may vary in space. Notice, that strictly speaking this setup is not realistic. It requires the presence of magnetic monopoles inside the torus. However, this does not affect the logic of our gedankenexperiment. We also assume that there may be the inhomogeneous electric potential of impurities, and the other sources of inhomogeneoity. We impose also the following constraint. The dependence of external fields (that cause the mentioned inhomogeneity) in the piece (II) of the torus repeats its dependence on coordinates inside the part (I).

{We represent potential $A_{\mu}$ as the sum: $A_{\mu}=A^{(m)}_{\mu} + A^{(e)}_{\mu}$. Here $A^{(m)}_{\mu}$ is gives rise to magnetic field and to potential of impurities.  $A^{(e)}_{\mu}$ corresponds to the external electric field. External electric field is weak, and uniform within each of the two mentioned above reasons. In region (I) for $y\in [0,L]$ the electric field is directed along the y-axis. Inside (II) for $y\in [-L,0]$ electric field has opposite direction. We represent the action (in Euclidean space - time) as follows
	\begin{eqnarray}
		S &=& \int d\tau \sum_{{\bf x,x'}}\Big[\bar{\psi}_{\bf x'}\Big(i(i \partial_{\tau} - A_3(i\tau,{\bf x}))\delta_{\bf x,x'} 
		- i{\cal H}_{\bf x,x'}\Big)\psi_{\bf x} {-}  \nonumber\\
		&& \alpha \bar{\psi}(\tau,{\bf x})\psi(\tau, {\bf x})
		\theta(y)V({\bf x-x'})\theta(y')\bar{\psi}(\tau,{\bf x'})\psi(\tau, {\bf x'})\Big] 
	\end{eqnarray}
Matrix ${\cal H}_{\bf x,x'}$ represents the one - particle Hamiltonian in the absence of interactions.
	$V$ is the interaction potential. We assume the instantaneous interactions. But the case when delay is present does not give anything new. For the particular case of Coulomb interactions it receives the form $V({\bf x})=1/|{\bf x}|=1/\sqrt{x_1^2+x_2^2}$,
	for ${\bf x}\not= {\bf 0}$.
	
Inside region (I) sufficiently far from the boundary with region (II) the momentum space action may be written in the form:
	$$S= \int  dp \bar{\psi}_{p}\hat{Q}(p,i\partial_p)\psi_{p}
{-}\alpha \int  dp dq dk \bar{\psi}_{p+q}\psi_{p}\tilde{V}({\bf q})\bar{\psi}_{k}\psi_{q+k},
	$$
	where $\hat{Q}(p,i\partial_p)$ is Dirac operator of the given non - homogeneous system
	\cite{Z2016_1}, while
	%
	$\tilde{V}({\bf q})$ is the Fourier transform of the interaction potential.
Fermion propagator with interaction corrections may be calculated as follows:
	\begin {eqnarray}\label{Green_a}
	G_{\alpha}(x,y) &=&
	G_{0}(x,y) + 
	\int G_{0}(x,z_1) \Sigma(z_1,z_2)G_{0}(z_2,y)dz_1 dz_2 +   \nonumber\\
	& & \int G_{0}(x,z_1) \Sigma(z_1,z_2)G_{0}(z_2,z_3)\Sigma(z_3,z_4)  
	G_{0}(z_4,y)dz_1 dz_2 dz_3 dz_4 +...
\end{eqnarray}
with
$\Sigma(z_1,z_2)=\add{-}\alpha G_{0}(z_1,z_2) \theta(y_1)V({\bf z}_1-{\bf z}_2)\theta(y_2)+O(\alpha^2)$,
 $z_i=({\bf z}_i,\tau_i)$, and $\bf z_i=(x_i,y_i)$.
Next we apply Wigner transformation:
\begin {eqnarray}\label{Green_b_Wigner}
G_{\alpha,W}(R,p) = G_{0,W}(R,p)  
+  G_{0,W}(R,p)\star \Sigma_W(R,p)\star G_{0,W}(R,p) + ...,  
\end{eqnarray}
In this expression $G_{0,W}(R,p)$ is solution of Groenewold equation $Q_{0,W}(R,p)\star G_{0, W}(R,p) = 1$. $\Sigma_W$ is the Wigner transformed self - energy $\Sigma$.

\subsection{Electric current expressed through the Green function with interaction corrections}

Let us expand $G_{\alpha,W}(R,p)$ in powers of $\alpha$: $G_{\alpha,W}= {\cal G}_0 + \alpha {\cal G}_1 +\alpha^2 {\cal G}_2 +... $
We also expand average electric current in powers of $\alpha$:
\begin {eqnarray} \label{current_3D_a}
\bar{J}^k(\alpha) &=& {-} \int \frac{d^2 R}{V} \int_p  Tr G_{\alpha,W}(R,p) \star 
\frac{\partial}{\partial p_k} Q_{0,W}(R,p)\nonumber\\
&=&  {-}\int \frac{d^2 R}{V} \int_p
Tr  \sum_{n=0}^{\infty} G_{0,W} 
(\star \Sigma_W\star G_{0,W})^n
\star \frac{\partial}{\partial p_k} Q_{0,W}(R,p) 
\end{eqnarray}
Here we denote by $R$ the spatial components of coordinates. The dependence on imaginary time $\tau$ is absent in our expressions. Therefore, variable $ x = (R,\tau)$ is reduced to $R$. Besides, we denote
$$
\int_p = \int \frac{d^3 p}{(2\pi)^3}
$$
Correspondingly, $\Sigma_W = \alpha \Sigma_{1,W} +\alpha^2 \Sigma_{2,W} +... $ and
$\bar{J}^{\mu}= \bar{J}^{\mu}_0 + \alpha \bar{J}^{\mu}_1 +\alpha^2 \bar{J}^{\mu}_2 +... $,
where
$$
\bar{J}^{k}_0= \int \frac{d^2 R}{V} \int_p Tr G_{0,W}\star \frac{\partial}{\partial p_k} Q_{0,W}
$$
Let us define the following quantity that is obtained when in the expression for the average current
$\partial_{p_k} Q_0$ is replaced by the  renormalized velocity:
\begin{eqnarray}\label{tildeI}
\tilde{J}^k(\alpha) = {-}\int \frac{d^2 R}{V} \int_p Tr G_{\alpha,W}(R,p) \star  
\frac{\partial}{\partial p_k}  Q_{\alpha,W}(R,p).
\end{eqnarray}
Here $G_{\alpha,W}(R,p)$ satisfies equation
\begin{equation}
G_{\alpha,W}(R,p)\star Q_{\alpha,W}(R,p) = 1 \label{groenewold}
\end{equation}
Since Eq. (\ref{tildeI}) is a topological invariant, its value cannot be changed, at least perturbatively, if $\alpha$ is increased smoothly from zero until the given value. Therefore, we have
$$
 \tilde{J}_k(\alpha) = \bar{J}_k(0)
$$
and
$$
 \bar{J}_k(\alpha) - \tilde{J}_k(\alpha) = \bar{J}_k(\alpha)-\bar{J}_k(0)
$$
We come to the conclusion that  $\tilde{J}_k(\alpha) = \bar{J}_k(\alpha)$ (i.e. the electric current may be calculated using renormalized velocity $\frac{\partial}{\partial p_k}  Q_{\alpha,W}(R,p)$) if there are no perturbative corrections to the total electric current. It appears that those corrections are indeed absent.

\subsection{Non - renormalization of Hall conductance by interactions}

\subsubsection{First order}

Electric current without interactions may be calculated as
$$
\bar{J}^k_0 =  {-}\int \frac{d^2 R}{V} \int_p  Tr G_{0,W}(R,p) \star \frac{\partial}{\partial p_k} Q_{0,W}(R,p)
$$
In \cite{ZZ2019_2} the proof is given that this expression does not receive corrections from interactions. In the first order we obtain:
\begin {eqnarray}\label{current_1st}
\bar{J}^k_1 &=&{+}\int \frac{d^2 R}{V} \int_{p,q} Tr  \Big( G_{0,W}(R,p-q)                   
D_W(R,q)\Big) \frac{\partial}{\partial p_k} G_{0,W}(R,p)
\end{eqnarray}
where $D_W$ is Wigner transform of
\begin {eqnarray}\label{Sigma_1}
D(z_1,z_2)&=&\add{-}\alpha \theta(y_1)V({\bf z}_1-{\bf z}_2)\theta(y_2),\label{DV}
\end{eqnarray}
Using integration by parts we arrive at  $\bar{J}_1^k = -\bar{J}^k_1$, which results in $\bar{J}_1^k = 0$. 

\begin{figure}[h]
	\centering  %
	\includegraphics[width=0.3\linewidth]{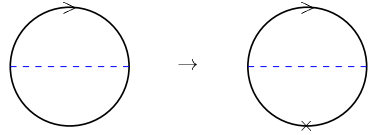}  %
	\caption{Example of "Progenitor" and the corresponding self-energy diagram. This is the case of the first order correction to electric current, when there is only one progenitor diagram.}  %
	\label{fig_cut-glue_2}   %
\end{figure}

\subsubsection{Second order}

{The second order contribution to electric current reads
\begin{eqnarray}\label{current_i2}
\bar{J}^k_2 = \new{+}\int \frac{d^2 R }{V}\int_p Tr \Sigma_{2,W} \star\partial_{p_k} G_{0,W}\new{+} 
\int \frac{d^2 R}{V} \int_p Tr \Sigma_{1,W} \star G_{0,W} \star \Sigma_{1,W}\star
\partial_{p_k} G_{0,W}\nonumber
\end{eqnarray}
which in rainbow approximation is given by
\begin{eqnarray}\label{current_i2}
&& \bar{J}^{k}_2 \approx   \nonumber\\
&&\add{+} \int  \frac{d^2 R }{V}  \int_{p,k,q} \,Tr \Big[G_{0,W}(R,p-k)
\star G_{0,W}(R,p-k-q)D_W(R,q) \star 
G_{0,W}(R,p-k)\Big] D_W(R,k)\star \partial_{p_k}G_{0,W}(R,p)  \nonumber\\
&&\add{+} \int  \frac{d^2 R }{V}  \int_{p,k,q} \,Tr G_{0,W}(R,p-q)D_W(R,q)
\star G_{0,W}(R,p)\star G_{0,W}(R,p-k)D_W(R,k)\star   %
\partial_{p_k}G_{0,W}(R,p) \nonumber\\
\end{eqnarray}
One can represent this result as
\begin{eqnarray}\label{current_i2}
I^{k}_2 &\approx & \add{+}\int  \frac{d^2 R }{V}\int_{p,k,q} \,Tr \Big[G_{0,W}(R,p-k) \star 
G_{0,W}(R,p-k-q)D_W(R,q)\star  
G_{0,W}(R,p-k)\Big]\star \nonumber\\
&& \quad\quad   D_W(R,k)\partial_{p_k}G_{0,W}(R,p)  \nonumber\\
&& {+}\int \frac{d^2 R}{V}\int_{\rvv{p,k,q}} \,Tr G_{0,W}(R,p-q)D_W(R,q)\star   
G_{0,W}(R,p)\star G_{0,W}(R,p-k)D_W(R,k)
\star \partial_{p_k}G_{0,W}(R,p)\nonumber\\
&=& {+}\frac{1}{2}\int  \frac{d^2 R }{V}\int_{p,k,q}\,\partial_{p_k} \,Tr \Big[G_{0,W}(R,p-k)\star  
G_{0,W}(R,p-k-q)D_W(R,q)\star 
G_{0,W}(R,p-k)\Big]\star \nonumber\\
&& \quad\quad  D_W(R,k)G_{0,W}(R,p), \nonumber
\end{eqnarray}
Since the last expression is the integral of total derivative, it is equal to zero. We encounter here the situation, when the  diagram for the electric current is obtained from the vacuum diagram via insertion of the derivative inside the integral. The latter vacuum diagram is called "progenitor". The progenitor also exists in the more simple first order approximation considered above, and it is  represented on the left - hand side of  Fig. \ref{fig_cut-glue_2}, while the   corresponding contribution to  electric current is represented on the right hand side of the same figure. }

The two loop diagrams  give
\begin{eqnarray}\label{current_i2}
I^{k(cross)}_2 &=& {+}\int \frac{ d^2 R}{V} \int_{p,k,q}\,Tr \Big[G_{0,W}(R,p-k)\circ_{2}\star  
G_{0,W}(R,p-k-q)\star \ _{1}\circ G_{0,W}(R,p-q)\star 
\partial_{p_k}G_{0,W}(R,p) \Big]\nonumber\\
& & D_{W(1)}(R,k)D_{W(2)}(R,q)   \nonumber\\
&=&  {+}\frac{1}{4}\int \frac{ d^2 R}{V} \int_{p,k,q} \partial_{p_k} \,Tr \Big[G_{0,W}(R,p-k)\circ_{2}\star  
G_{0,W}(R,p-k-q)\star \ _{1}\circ  G_{0,W}(R,p-q)\star  
G_{0,W}(R,p) \Big] \nonumber\\
& & D_{W(1)}(R,k) D_{W(2)}(R,q)   \nonumber
\end{eqnarray}
The last expression also vanishes. In this expression we use the following notations. Operator
$\star=e^{i\overleftarrow{\partial}_R\overrightarrow{\partial}_p/2-i\overleftarrow{\partial}_p\overrightarrow{\partial}_R/2}$
acts only on $G$ and does not act on $D$, while
$\circ_{i} = e^{-i\overleftarrow{\partial}_p\overrightarrow{\partial}_R/2}$
contains derivatives with respect to $p$ and $R$. The derivatives with the right arrow act on $D_{W(i)}$, while the derivatives with the left arrow act on the  fermion propagator standing left to this symbol.
In operator $_{i}\circ=e^{i\overleftarrow{\partial}_R\overrightarrow{\partial}_p/2}$
the derivatives with the right arrow
act on the expression following this symbol while the derivatives with the left arrow act on $D_{W(i)}$.
Notice that $D_{W(i)}$ does not contain $p$. Therefore, operator $\circ$ does not affect $G$. 

 This way we prove that all second order contributions to electric current vanish. The consideration of the radiative corrections of arbitrary order is, in principle, similar, but more involved. The complete proof that those contributions vanish is given in \cite{ZZ2021}.

\section{Application of Wigner - Weyl calculus to the QHE in non - homogeneous systems out of equilibrium}
\label{SectKeldysh}

In \cite{BFLZZ2021} the approach of  \cite{Shitade, Sugimoto,Sugimoto2006,Sugimoto2007,Sugimoto2008} to the incorporation of Wigner - Weyl calculus to Keldysh technique has been developed further. Here we briefly summarize the obtained results.
Keldysh Green function is matrix
\begin{eqnarray}
	\hat{ G}(t,x|t^\prime,x^\prime)
	= -i \left(\begin{array}{cc}\langle T \Phi(t,x) \Phi^+(t^\prime,x^\prime)\rangle & -\langle  \Phi^+(t^\prime,x^\prime) \Phi(t,x)\rangle\\ \langle  \Phi(t,x) \Phi^+(t^\prime,x^\prime)\rangle & \langle \tilde{T} \Phi(t,x) \Phi^+(t^\prime,x^\prime)\rangle \end{array} \right).
	\label{KelG_S}
\end{eqnarray}
In this expression the Heisenberg fermionic field operator $\Phi$ is a function of both spatial coordinates $x$ and time $t$. Symbol $T$ means the time ordering while $\tilde{T}$ is the anti - time ordering. By $\langle ... \rangle$ the average with respect to the initial state is denoted.

By large latin letters we denote the $D$ dimensional space - time vectors. Operator $\bf G$ corresponding to the above introduced Green function is given by:
$\hat{G}(X_1,X_2) = \langle X_1 | \hat{\bf G} | X_2 \rangle $. $\bf Q$ is an operator  inverse to $\bf G$.

Matrix elements  $A(X_1,X_2) = \langle X_1 | \hat{A} | X_2 \rangle $  of any operator $\hat{A}$  give rise to the Weyl symbol of $\hat A$:
\be
A_W(X|P)=\int d^{D} Y\, e^{\ii Y^\mu P_\mu }A(X+Y/2,X-Y/2),\quad \mu =0,1,...,D-1
\label{WignerTr_S}
\ee
Weyl symbols of $\hat{G}$ and $\hat{Q}$  obey the Groenewold equation $Q \star G = 1$ with the Moyal product
\begin{equation}
	\left(A\star B\right)_W(X|P) = A_W(X|P)\,e^{\rv{-}\ii(\overleftarrow{\partial}_{X^{\mu}}\overrightarrow{\partial}_{P_{\mu}}-\overleftarrow{\partial}_{P_{\mu}}\overrightarrow{\partial}_{X^{\mu}})/2}B_W(X|P).
\end{equation}
Application of this technique to the QHE of two - dimensional systems allows to obtain  following results \cite{BFLZZ2021}:

	The DC conductivity of the two - dimensional {\it non - interacting} systems is given by
		\begin{equation}
		\sigma^{ij} =  {\frac{1}{4}} \int \frac{d^{\rv{3}}P}{(2\pi)^{\rv{3}} } \tr\left(\gamma^< \partial_{P_{i}}\hat{Q}_W  \left[\hat{G}_W \star \partial_{\rv{P_{[0}}}\hat{Q}_W  \star \partial_{\rv{P_{j]}}}\hat{G}_W  \right]\right) +{\rm c.c.}\label{MAIN_S}
	\end{equation}
	Here trace is taken over Keldysh components as well as over the internal indices, and
	$$
		\gamma^< = \left(\begin{array}{cc} 0 & 0 \\ 1 & 0 \end{array} \right).
		$$

	It was shown that the above expression for the conductivity averaged over the system area is reduced to the topological formula Eqs. (\ref{JQHE}), (\ref{JQHE2}).
 The intermediate expression through the real - time Green functions obtained in our derivation reproduces the one of \cite{Mokrousov}.

Interaction corrections to the QHE conductivity were calculated, and it was shown that already in the one loop order those corrections do not vanish in case of equilibrium at finite temperature, and also out of equilibrium.

\section{Conclusions}

\label{SectConcl}

To conclude, in this paper we review the results obtained by the group working in Ariel University. We summarize here the results reported previously in a series of papers \cite{ZW2019,ZZ2019_2,FSWZZ2020,ZZ2021,SZ2020,BLZ2021,BFLZZ2021,ZZ2019_FeynRule} (see also references therein to the other papers of the group). In these works the non - dissipative transport phenomena have been investigated using the technique of Wigner - Weyl calculus. The obtained results concern the chiral magnetic effect (CME), the chiral separation effect (CSE) and the quantum Hall effect (QHE). The considered systems are essentially non - homogeneous. This is why the use of the Wigner - Weyl calculus for obtaining the topological expressions seems to us inevitable.

Below we summarize the main obtained results

\begin{enumerate}

\item{}

 The approximate Wigner-Weyl calculus for the lattice models has been developed. It allows to deal with the systems in the presence of weak inhomogeneity. Weakness of inhomogeneity means that physical quantities almost do not vary  at the distance of the order of the lattice spacing. Within this calculus we develop Wigner - Weyl field theory in equilibrium \cite{ZW2019,FSWZZ2020}, and its extension to quantum kinetic theory \cite{BFLZZ2021}. In this theory currents and conductivities are expressed through the Wigner transformed fermion propagators and Weyl symbols of operators.

\item{}

It is shown that the CME conductivity is absent in true equilibrium both at zero and at finite temperatures \cite{BFLZZ2021}. Weak inhomogeneity cannot affect this conclusion. Inter - electron interactions also do not affect this conclusion, at least, in one - loop order. Notice that we did not consider here the zero frequency limit of non - equilibrium CME conductivity. The latter limit is to be the subject of a separate study and it might be different from the case of the true equilibrium, when the system is not subject to the time dependent chiral chemical potential.

\item{}

We have shown that the radiative corrections to the QHE conductivity in the two - dimensional systems are absent. We have shown this first in the first and the second orders of perturbation theory \cite{ZZ2019_2}, and later extended this proof to the third order and to the higher orders of perturbation theory \cite{ZZ2021}. Besides, we demonstrate that the QHE conductivity in the presence of interactions may be calculated using the same expression as the QHE conductivity of the non - interacting systems, in which the Green function is replaced by the complete interacting one.

\item{}

The CSE is considered for the non - homogeneous systems with chiral fermions \cite{SZ2020}. We assume that in these systems the effective low energy theory obeys chiral symmetry. On the language of Green functions it is expressed as follows: matrix $\gamma^5$ commutes or anti - commutes with the Wigner transformed Green function, and with the Weyl symbol of lattice Dirac operator (the latter is inverse to the Green function). This requirement is imposed on piece of phase space with any values of coordinates, but for the region of momenta that belongs to a vicinity of the Fermi surface. For such systems the CSE conductivity is given by the integral over coordinates $x$ and for each value of $x$ over the $3D$ hypersurface surrounding the analogue of the Fermi surface. The latter is understood as the points in momentum space, where expression standing in the mentioned integral has poles.

\item

In the framework of Keldysh theory we consider the QHE conductivity and derive its useful representation through the Wigner transformed Keldysh propagators \cite{BFLZZ2021}. As it should, this representation is reduced to the topological one of \cite{ZW2019} when the system approaches equilibrium with vanishing temperature. At the same time out of equilibrium and in thermal equilibrium with finite temperature the QHE conductivity is not robust to the smooth modification of the system. Besides, our consideration of interaction corrections demonstrates that the radiative corrections to the QHE conductivity do not vanish out of equilibrium as well as in thermal equilibrium with finite temperature.

\end{enumerate}

The results reviewed in the present paper have been obtained in collaboration with C.Banerjee, I.Fialkovsky, M.Lewkowicz, M.Suleymanov, X.Wu, C.X.Zhang. The author is grateful to all former and present members of the group for this collaboration and for the important contributions to the obtained results. Besides, the author kindly acknowledges numerous discussions with G.E. Volovik and Z.V.Khaidukov.





	


\bibliography{wigner3,cross-ref,buotbibl,bibcse}

\end{document}